\def\be{\begin{equation}}
\def\ee{\end{equation}}
\def\bea{\begin{eqnarray}}
\def\eea{\end{eqnarray}}
\def\bse{\begin{subequations}}
\def\ese{\end{subequations}}

\documentclass[prb,twocolumn,showpacs,amsmath,amssymb,eqsecnum]{revtex4}
\usepackage{graphicx}
\usepackage{dcolumn}
\usepackage{bm}
\begin{document}
\title{Reply to ``Comment on `Fluctuation-induced first-order transition in $p\,$-wave superconductors' ''
}
\author{Qi Li,$^1$ D. Belitz,$^2$ and John Toner$^2$}
\affiliation{$^1$ National Institute of Neurological Disorders and Stroke,
                           National Institute of Health, Bethesda, MD 20892\\
             $^2$ Department of Physics and Institute of Theoretical Science,\\
                  University of Oregon, Eugene, OR 97403
         }
\vskip -20mm
\date{\today}


\maketitle

We regret the inaccuracies, and especially the omission of pertinent references, in our
paper,\cite{Li_Belitz_Toner_2009} which are pointed out in D.I. Uzunov's Comment,\cite{Uzunov_2012}
the technical aspects of which we agree with. Accordingly, we would like to make the following four
corrections to Ref.\ \onlinecite{Li_Belitz_Toner_2009}.
\bigskip\par\noindent
(1) The end of Sec. III.A, starting after Eq. (3.3) and including Fig. 1, should read

\medskip
We now need to distinguish between two cases.

\smallskip
{\it Case 1:} $v>0$. The free energy is minimized by either ${\hat n} = \pm{\hat m}$ and arbitrary 
$\phi$, or by $\phi = k\pi/2$, $k = 0,\pm1,\ldots$ and arbitrary 
${\hat n}\cdot{\hat m}$,\cite{Volovik_Gorkov_1984, Volovik_Gorkov_1985, Ueda_Rice_1985,
Blagoeva_et_al_1990}
The condition $u>0$ must be fulfilled for the system to be stable.

\smallskip
{\it Case 2: $v < 0$.} The free energy is minimized by ${\hat n} \perp {\hat
m}$ and $\phi = \pi/4$, and $\psi_0 = -t/2(u + v)$. The condition $u + v
> 0$ must be satisfied for the system to be stable.

\smallskip
Phases with ${\bm\psi}\times{\bm\psi}^* = 0$ and ${\bm\psi}\times{\bm\psi}^* \neq 0$ are sometimes 
referred to as unitary and nonunitary phases, respectively. In all cases, mean-field theory predicts a
continuous transition from the disordered phase to an ordered phase at $t=0$. The mean-field phase
diagram in the $u$-$v$-plane is shown in Fig. 1.

\begin{figure}[t,b,h]
\includegraphics[width=4.0cm]{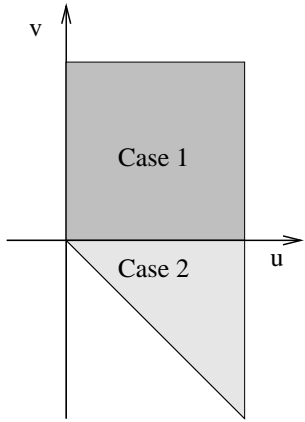}
\caption{Mean-field phase diagram of a $p\,$-wave superconductor as described
 by Eq.\ (2.1). See the text for additional information.}
\label{fig:1}
\end{figure}

\bigskip\par\noindent
(2) The text after Eq. (3.5) should read

\medskip
Here $w \propto (\mu q^2)^{3/2}$ is a positive coupling constant whose presence drives the transition
into any of the ordered phases first order. In addition, the parameters $t$ and $u$ acquire dependences
on $\mu q^2$, as is the case for s-wave superconductors.\cite{Halperin_Lubensky_Ma_1974}

\bigskip\par\noindent
(3) Equations (3.7), the sentence preceding them, and the paragraph following them, should read

\medskip
Defining ${\tilde u} = u+v$, and ${\tilde v} = -v$, we find the RG recursion relations
\be
\frac{dt}{dl} = (2-\eta)\,t + 3\,c\,q^2\mu + \frac{(n+2){\tilde u} + 4{\tilde v}}{t+c}\ ,\hskip 20pt
 \tag{3.7a}
\ee
\vskip -20pt
\be
\frac{d{\tilde u}}{dl} = (\epsilon-2\eta)\,{\tilde u} - \frac{(n+8){\tilde u}^2 + 8{\tilde u}{\tilde v} 
   + 8{\tilde v}^2}{(t+c)^2} - 3\,c^{2}q^{4}\mu^{2}, \nonumber\\
\tag{3.7b}
\ee
\vskip -20pt
\be
\frac{d{\tilde v}}{dl} = (\epsilon-2\eta)\,{\tilde v} - \frac{n{\tilde v}^2 + 12{\tilde u}{\tilde v}}{(t+c)^2}\ ,
\hskip 65pt
\tag{3.7c}
\ee
\vskip -20pt
\be
\frac{dc}{dl} = -\eta\,c - 3\frac{c^2 q^2 \mu}{t+c}\ ,\hskip 105pt
\tag{3.7d}
\ee
\vskip -20pt
\be
\frac{d\mu}{dl} = \eta_A\,\mu - \frac{n}{6}\, \frac{(3t+c)c^3 q^2 \mu^2}{(t+c)^4}\ ,\hskip 64pt
\tag{3.7e}
\ee
\vskip -20pt
\be
\frac{dq}{dl} = \frac{1}{2}\,(\epsilon - \eta_A)\,q\ .\hskip 120pt
\tag{3.7f}
\ee
Here $l = \ln b$ with $b$ the length rescaling parameter, we have redefined
$4\pi\mu \to \mu$, and we have absorbed a common geometric factor in the
coupling constants $u$, $v$, and $\mu$. These flow equations were first derived 
by Millev and Uzunov,\cite{MIllev_Uzunov_1990} and later generalized to include 
effects of quenched disorder.\cite{Busiello_et_al_1991} For $v=0$, they reduce to those of 
Ref.\ \onlinecite{Halperin_Lubensky_Ma_1974}, as they should.

\bigskip\par\noindent
(4) In the remainder of the paper, $u^*$ and $u_0^*$ should be interpreted as ${\tilde u}^*$, and
$v^*$ and $v_0^*$ as ${\tilde v}^*$. 


\end{document}